# Strongly enhanced light-matter coupling of a monolayer WS$_2$ from a bound state in the continuum

E. Maggiolini[1,2]*, L. Polimeno[1]*, F. Todisco [1], A. Di Renzo[1,3], M. De Giorgi[1], V. Ardizzone[1], R. Mastria[1], A. Cannavale[1,4], M. Pugliese[1], V. Maiorano[1], G. Gigli[1,3], D. Gerace[2], D. Sanvitto[1]†, D. Ballarini[1]†

[1]*CNR NANOTEC, Institute of Nanotechnology, Via Monteroni, Lecce 73100, Italy*

[2]*Dipartimento di Fisica, Universita` di Pavia, 27100 Pavia, Italy*

[3]*Dipartimento di Matematica e Fisica E. De Giorgi, Universita` del Salento, Campus Ecotekne, Via Monteroni, Lecce 73100, Italy*

[4]*Department of Civil Engineering Sciences and Architecture, Polytechnic University of Bari, Bari, Italy*

† *Corresponding author: dario.ballarini@nanotec.cnr.it, daniele.sanvitto@nanotec.cnr.it*

**Optical bound states in the continuum (BIC) allow to totally prevent a photonic mode from radiating into free space along a given spatial direction. Polariton excitations derived from the strong radiation-matter interaction of a BIC with an excitonic resonance inherit an ultra-long radiative lifetime and significant nonlinearities due to their hybrid nature. However, maximizing the light-matter interaction in these structures remains challenging, especially with 2D semiconductors, thus preventing the observation of room temperature nonlinearities of BIC polaritons.**

---

*These authors contributed equally: E. Maggiolini, L. Polimeno

**Here we show a strong light-matter interaction enhancement at room temperature by coupling monolayer WS$_2$ excitons to a BIC, while optimizing for the electric field strength at the monolayer position through Bloch surface wave confinement. By acting on the grating geometry, the coupling with the active material is maximized in an open and flexible architecture, allowing to achieve a 100 meV photonic bandgap with the BIC in a local energy minimum and a record 70 meV Rabi splitting. Our novel architecture provides large room temperature optical nonlinearities, thus paving the way to tunable BIC-based polariton devices with topologically-protected robustness to fabrication imperfections.**

**Introduction**

Bound states in the continuum (BICs) are solutions of a wave equation that lie within the range of continuum energy states, but without any coupling to far-field radiation [1,2]. As a consequence, BICs are characterized by extremely large quality factors (Q), only limited by material losses. Moreover, in some cases, they can be featured by a topological protection, resulting in high robustness to external perturbations and fabrication imperfections [3]. Despite being conceived as potential engineering tools to confine electrons - and as such difficult to realize - they have recently been extended in many wave phenomena in periodic lattices, spanning from the acoustic[4–6] to the electromagnetic[7,8] waves domain.

Thanks to the advances in material processing, nanophotonics has rapidly emerged as a fruitful platform to develop structures sustaining optical BICs, which have been observed in several architectures including photonic crystals [8], waveguides [9] and metasurfaces [10]. Notably, the large Q factor of these states makes nanophotonic BICs extremely appealing for enhancing light-matter interactions when integrated with optically active materials [11,12]. From this point of view, strongly coupled BICs with different materials have been reported [13,14], giving rise to unique optical properties of the so called exciton-polaritons. These bosonic quasi-particles arising from the strong coupling between excitons and photons [15], acquire hybrid properties inherited from their bare components, among which one of the most pursued for is the strongly interacting character [16,17]. In particular, one of the most interesting characteristic of polariton BICs is their lifetime which is largely increased compared to the standard exciton-polaritons [13,14]. This is a crucial step to bring polariton physics to room temperature operation, paving the way for the realization of ultrafast nonlinear electro-optical applications, such as sensing [18], low-threshold lasing [19], information processing and logic operations [20]. A first milestone in this direction has been recently achieved with the demonstration of polariton condensation from a BIC state by using patterned GaAs waveguides at cryogenic temperature [21]. However, with the aim of achieving room temperature nonlinear operation, exciton-polaritons in semiconductors with larger exciton binding energy are required. In this respect, atomically thin transition metal dichalcogenides (TMDs) represent an ideal active medium for realizing room-temperature polariton systems, thanks to a stable excitonic resonance arising from their 2D confinement and reduced dielectric screening [22]. However, coupling photonic BIC states with single monolayers poses further issues, since the BIC modes are strongly confined withing waveguides while the 1D semiconductor can only be deposited at their surface, rendering the coupling with the BIC field quite weak. In the unique experimental realisation so far, a TMD monolayer is transferred on top of the patterned PhC slab, limiting the overlap between the electric field profile and the exciton wave function, therefore resulting in a vacuum Rabi splitting comparable with the exciton linewidth [23].

An alternative solution is provided by Bloch Surface Waves (BSWs), evanescent electromagnetic modes propagating at the interface between a Distributed Bragg Reflector (DBR) and the surrounding medium [24]. In a BSW, the electric field enhancement is localized at the surface of the dielectric substrate, allowing narrow resonances that couple strongly with thin films depositedon the top of the DBR, such as organic materials [25] or TMDs [26]. Moreover, in the context of BIC design, BSWs allow for a remarkably flexible and sensitive tuning of their spatial field pro- files, thanks to their ability to well support modes both in gaps and fills of the pattern and with a superior implementability from the fabrication point of view [27].

In this work, we report the systematic engineering and observation of a topologically-protected BIC state in an energy minimum of the counter-propagating BSW polaritons (BSWP). We achieve a record strong coupling of 70 meV between a tungsten disulfide ($WS_2$) monolayer and the photonic BIC mode of a patterned BSW, following from a multiparametric metaheuristic optimization approach of the whole system and the fabrication-wise integration of the active material beneath the patterned structure .
We conceptually show - see sections A, B, C of the SI and Fig. S1-4 for a thorough treatment - and experimentally demonstrate that our design allows to simultaneously fulfill several stringent requirements, i.e. 1) maximizing light-matter interaction between photonic BIC and TMD exciton, 2) BIC state placed at the local minimum of the energy dispersion, rather than the maximum of the two bands, as commonly realised, 3) large spectral decoupling from the lossy mode at lower energy, 4) tolerant and flexible fabrication design. Moreover, by resonantly pumping the polariton dispersion, we measure a polariton blueshift of $\sim$ 15 meV at high pump fluences, demonstrating that our system sustains strong polariton nonlinearities at room temperature.

Combining the enhanced Coulomb interactions in a $WS_2$ monolayer and an easy tunability of the BIC resonance in the BSWP dispersion through our design approach, we theoretically propose and experimentally show a unique platform to manipulate the topological properties in a strongly interacting system, paving the way for the realization of a novel class of optical devices working at room temperature.

**Results**

Our sample architecture is sketched in Fig. 1A and consists of a layered $SiO_2$-$TiO_2$ DBR fabricated by electron-beam evaporation on a thin glass substrate. The structure is designed with a stopband centered at 1476 meV, supporting a Transverse Electric (TE) propagating BSW beyond the light cone with energy $E_{BSW} \simeq$ 2 eV at k=11.3 µm$^{-1}$ (Fig. S6A). A monolayer of $WS_2$ is then transferred on top of the DBR using a dry transfer method with commercial polydimethylsiloxane (PDMS). Finally, a 20 nm-thick $SiO_2$ film is evaporated on the whole sample as a protective layer. The capped $WS_2$ monolayer supports a sharp excitonic resonance at 2016 meV (see Fig. S5), crossing the BSW dispersion beyond the light cone. We measure the reflection spectrum of the coupled $WS_2$-BSW system in the in-plane momentum space by using an oil immersion objective, showing the typical anticrossing behaviour of the strong coupling regime (Fig. S6B). The grating is fabricated on the top of the SiO$_2$ encapsulation layer by electron beam lithography of a polymethyl-methacrylate (PMMA) thin film (see Methods for fabrication details). The reflectivity spectrum of the fabricated device, measured in Fourier space in the region around k=0, is shown in Fig. 1B. The obtained BSW-polariton dispersion is then used to optimise the design of a non-trivial 1D PhC with multiple harmonics in the Fourier spectrum (schematically shown in Fig. 1C), which is able to both fold the BSW-polariton dispersion and to mirror it back at zero momentum, while at the same time maximizing the light-matter interaction.

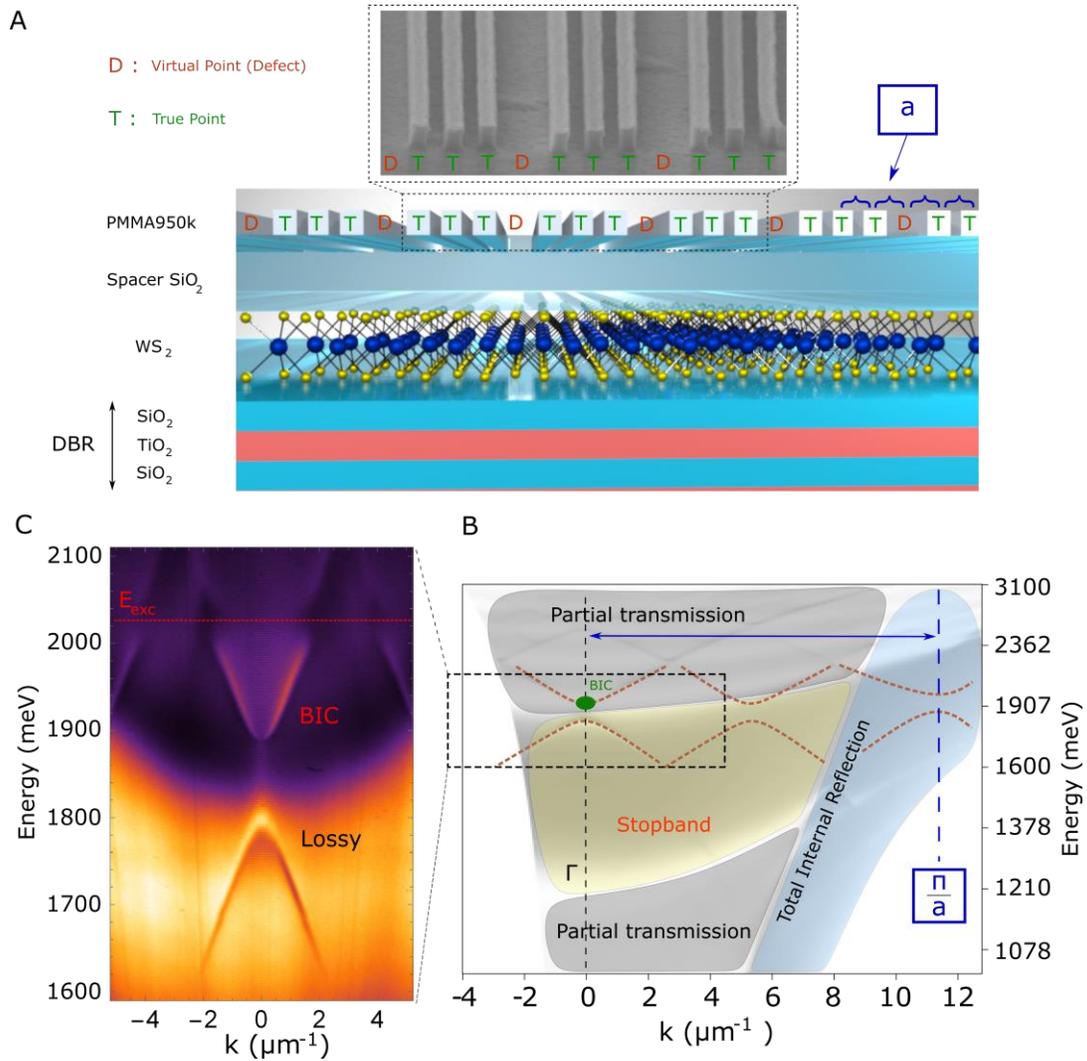

Figure 1: A) Sketch of the device. The PMMA grating is fabricated on the top of a DBR-WS$_2$-SiO$_2$ structure with a defected subwavelength structure - pillars marked by D are systematically missing -, leading to harmonics at K$_G$/4, K$_G$/2, 3K$_G$/4 and K$_G$ (see section A of the SI), where $K_G = \frac{\pi}{a}$ and "a" is the subwavelength periodicity. Inset: SEM image illustrating the high-aspect-ratio fabrication process, where each 3-element grating unit is made up of 2 lateral pillars of ∼ 177 nm width and a central pillar of ∼ 118 nm width. The scale bar is 500 nm. B) Experimental reflectivity in k-space of the sample, zooming the region close to $k \sim 0$. C) RCWA simulation of the grating effect on the BSW supported by the DBR. The defected grating harmonic K$_G$ induces a wide gap-opening at the bandedge, while the K$_G$/2 harmonic mirrors the dispersion to $k \sim 0$, leading to a true BIC at the Γ point.

**Design and optimization**

Among the possible optical BICs one can mold through a PhC patterning, here we focus on the specific case of the topologically-protected BIC occurring at the Γ point in the reciprocal lattice of a 1D PhC. These BICs - among the first to be studied in photonics[2] - are antisymmetric modes with respect to the real-space symmetry axes of the lattice; as such, they cannot couple to the continuum of radiation in free space, which is symmetric at the Γ point. Compared, for instance, to non-simmetry-locked BICs obtained by parameter tuning (so called accidental BICs), symmetry-protected Γ-BICs hold their topologically-protected nature even with the $σ_z$ simmetry broken - z is the stack growth direction -, which is exactly the case for our BSW system. Moreover, their pinning by construction to the Γ point leaves the energy tuning as the only fabrication-sensitive degree of freedom; their overall design robustness to fabrication imperfections is thus remarkable. However, while Γ-BICs can be realised by tuning the filling factor of a simple periodic grating, here a more sophisticated approach is essential to the simultaneous enforcement of multiple - in principle independent targets, i.e., maximizing the light-matter coupling, ensuring a BIC energy valley at a specific detuning from the exciton and creating the conditions for a suppression of other modes allowed in the system with respect to the BIC-hosting branch. We proceed to lock the last two points by a wise choice of the general design concept, then meeting them with the upmost important target of the interaction enhancement, by a metaheuristic parametric optimization over the chosen kind of structure. Indeed, the amplitude of the gap for a subwavelength grating are irreproducible by a superwavelength grating because in the latter the bandfolding is due to higher order harmonics in the Fourier spectrum, while in the subwavelength case the folding is due to the strong first harmonic. However, to achieve an effective mirroring of the dispersion in $k=0$, additional components in the Fourier spectrum are needed. Here we modify the subwavelength grating structure by the subtraction of one pillar every two periods, obtaining a grating with both well defined and fabrication-robust folding and mirroring components in the Fourier spectrum, as schematically shown in Fig. 1C and in Fig. S1D of the SI. We note that both the lossy and BIC modes in principle increase in linewidth due to the mirroring to $k=0$, but while in the BIC state this effect does not manifest due to the symmetry-locked protection, the lossy mode acquires a strongly non-hermitian character, which is possible to engineer to result decoupled from the active material (see SI for more details). A key point in our design is the choice of an odd number of pillars in each subwavelength unit, allowing us to quasi-deterministically force the antisymmetric BIC-like mode as the upper dispersion branch. Indeed, the antisymmetric BIC-like mode must exhibit a node in its electric field at each symmetry axis of the structure: the central pillar of each subgroup is selectively probed by the antinode of the symmetric lossy mode. Therefore, the defective grating increases the symmetric (lossy) mode effective index, while it leaves mostly unperturbed the antisymmetric (BIC-like) mode. Finally, a metaheuristic approach extensively described in section B of the SI, combines heuristics - i.e. physical considerations in the setting of Finite Difference Time Domain (FDTD) simulations - and an evolutive algorithm to provide a flexible and quick-convergent procedure for the co-optimisation of the grating and stack parameters. This approach allows us to maximise the light-matter coupling with a full control over the BIC state as a global parametric optimisation, jointly attaining multiple results impossible with single-variable parameter sweeps. In section C of the SI we also show that our experimentally-demonstrated record Rabi splitting - see next section - is close from below to the one of an ultra strong coupling condition, beyond which paradoxical decoupling effects can even appear between light and matter. By qualitative and numerical considerations, in the same section of the SI we show that our optimization algorithm is robust and meaningful while optimizing the light-matter interaction strength from a weak-coupling condition up and across a strong coupling one and then almost up to the ultra strong coupling.

**Strong coupling and room temperature nonlinearities**

The energy vs in-plane momentum photoluminescence of the $WS_2$ monolayer after the fabrication of the grating is shown in Fig. 2A, off-resonant excited by using a continuous-wave 488 nm diode laser. The formation of BIC polaritons in the minimum of the energy dispersion at k=0 is clearly visible as a vanishing intensity for k→0, where the destructive interference of optical waves at k=0 induces the light trapping in the near field and vanishing far-field radiation. At the same time, the opening of a wide energy gap (∼ 100 meV) between the BIC and the lossy lower mode is also visible, effectively decoupling the exciton from the lossy mode. The experimental Q values (red square) and the polariton linewidths (black dots) in the BIC branch are reported as a function of the in-plane momentum in Fig. 2B. Decreasing the in-plane momentum, the polariton linewidth narrows and results in a burst of the Q factor approaching k=0.

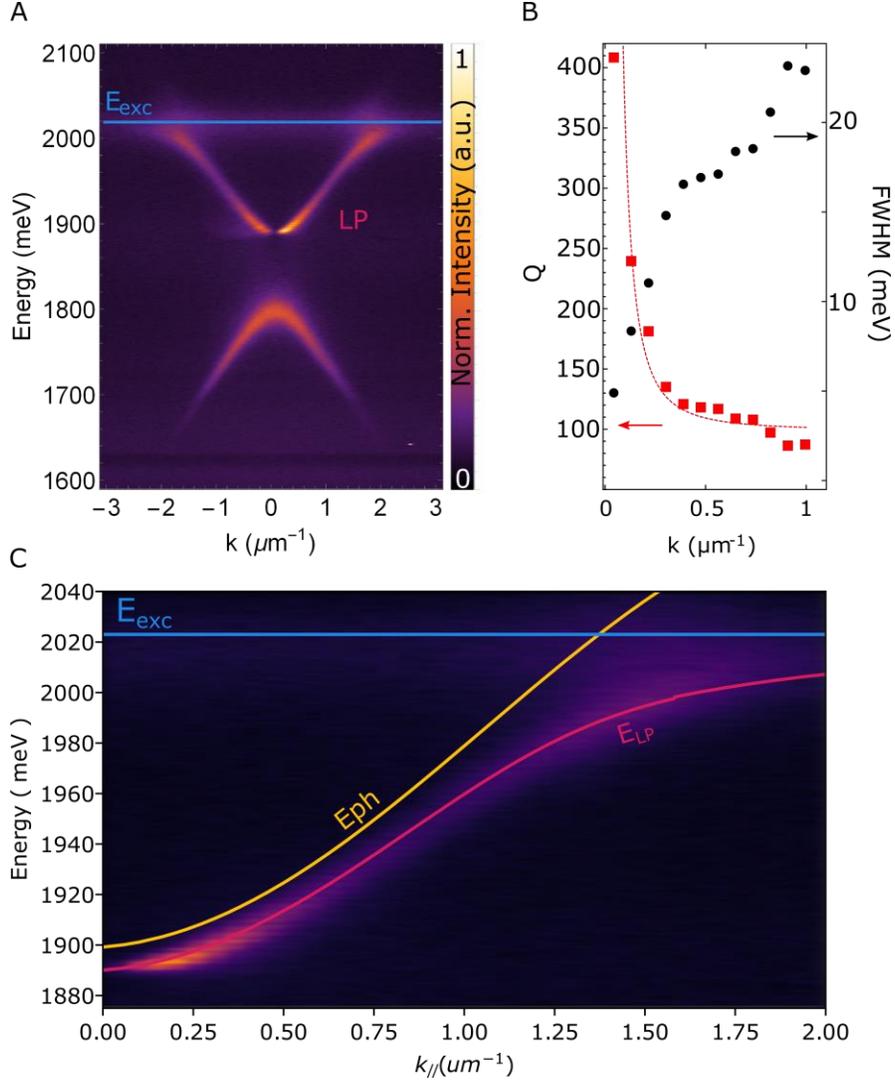

Figure 2: A) Energy vs in-plane momentum photoluminescence of WS$_2$ monolayer, strongly coupled with the BIC mode generated by the grating. The red dashed line indicates the exciton resonance $E_{exc}$. B) Q factor values (Black dots) and polariton linewidths (red square) plotted as a function of the in-plane momentum. C) Fitting of the photoluminescence dispersion used to extract the Rabi splitting. The blue solid line represents the $E_{exc}$ resonance; in yellow the photonic dispersion $E_{ph}$ estrapolated from reflection measurements on a TMD-free area; in red the fitted polaritonic dispersion for a 70 meV Rabi splitting (see section E of the SI for additional details on the calculation method).

The anticrossing of the BIC branch with the monolayer exciton is shown in detail in Fig. 2C. Notably, we extract a Rabi splitting of $\Omega \sim$ 70 meV, that is higher than previous values of Rabi splitting obtained for TMDs in the strong coupling regime, both in planar microcavity [28–30] and in other photonic structures [23, 26]. This is obtained thanks to the precise control of the spatial enhancement of the BIC-like BSW mode in proximity of the monolayer. To investigate the presence of optical nonlinearities at room temperature (RT), we resonantly excite the polariton mode with a femtosecond pulsed laser close to the BIC state, at k=0.5 µm$^{-1}$. Figure 3A shows that the reflectivity spectrum of the laser changes as the incident power is increased. The low-fluence spectrum (black solid line) shifts as the incident fluence increases (light-blue solid line), recovering its initial lineshape when the incident fluence is lowered (red dashed line). This reversibility guarantees that the blueshift is not due to degradation of the material, but that it arises from polariton interactions. In the mean-field approximation, the polariton blueshift $\Delta E_{pol}$ is defined as $\Delta E_{pol} = g_{pol} \cdot n_{pol}$, where $n_{pol}$ is the polariton density, proportional to the excitation power, and $g_{pol}$ is the polariton nonlinearity. By fitting the data reported in Fig. 3A as a function of the incident fluence with a straight line (Fig. 3B), we estimated the polariton interaction constant at RT, $g_{pol} \sim 5.8 \cdot 10^{-4} \mu eV \, \mu m^2$. Knowing the polariton excitonic fraction of the mode, $X \sim 0.08$, we extract the excitonic interaction constant $g_{exc} \sim 6 \cdot 10^{-2} \mu eV \, \mu m^2$, in accordance with the $g_{exc}$ value found for WS$_2$ monolayer at room temperature [26]. The $g_{exc}$ extracted from our experiment neglects any loss in the material, overestimating the number of polaritons pumped in the system: larger interaction strengths should be expected on our platform.

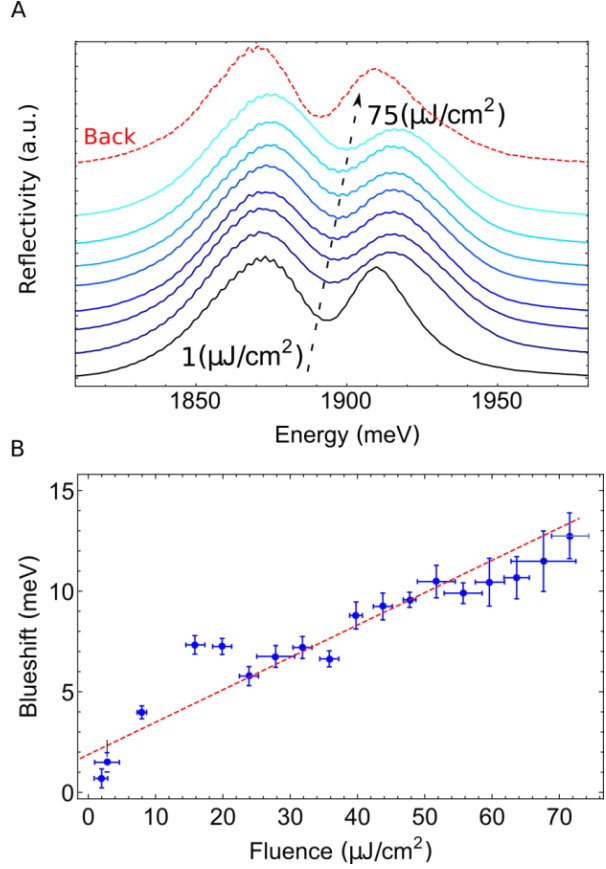

Figure 3: A) Reflection spectra of the pulsed laser on polariton mode at k $\sim$ 0.5 $\mu m^{-1}$, corresponding to different resonant excitation fluences. B) Energy blueshift of the polariton modes at k $\sim$ 0.5 $\mu m^{-1}$ as a function of the incident fluence.

**Discussion**

Our results show that the combination of BSWs, 2D materials and a non-trivial 1D PhC through a global optimization approach, allows to build large light-matter interactions in topologically-protected BIC-polaritons. Thanks to the record Rabi splitting of 70 meV, we can achieve large polariton nonlinearities close to a BIC state at room-temperature. Moreover, while BICs have been typically realized on energy maxima (lower band) of photonic crystal waveguide modes[12, 21, 23], we show that BICs at the minimum of the dispersion (upper band) can be formed with our design that optimises also the gap between the two bands and the light-matter interaction strength at or beyond the present state of the art. This is an interesting result, since having an energy valley well decoupled from other modes can provide a state with reduced leakages and accumulating population scattered from higher energy states ; improving on these two points while assuring an high light-matter interaction strength is clearly a crucial requirement for low-threshold lasing or low-threshold polariton condensation at room temperature in an open system.[31].

A further development made possible by our design approach is the possibility to turn the BIC energy minimum into an absolute minimum, within a polaritonic dispersion which emulates in all regards a single-mode microcavity, which thanks to the BIC properties has a long-lifetime valley. Indeed, through our defective grating design, we set the condition to achieve the enhancement of the non-hermitian nature of the non-BIC lower photonic branch, well beyond the enhanced losses due to the perfect symmetry matching to the continuum radiation at $k = 0$. Viewed in the framework of QNM (Quasi-Normal-Mode) theory[32], this introduces a greater degree of freedom in the numerical tuning of the dispersion characteristics, paving the way towards a strong and selective suppression of the light-matter coupling for the lower lossy mode.

Finally, we are able to reach the above-mentioned results having the 2D material successfully encapsulated in a glass stack and the Γ-BIC resonance spatially enhanced both at the level of the active material as well as open-air. The possibilities offered by a platform of this kind, with an active part well protected but yet able to strongly interact with the environment, go for instance from applications in bio-molecular sensing - through the embedding in microfluidic systems - to fundamental studies on dipolar interactions[33], Rydberg polaritons[34] and electrically-tuned non-neutral optical transitions[35] in TMDs - through the possible integration of transparent contact layers.

## 1 Methods

**Sample Fabrication** The DBR is formed by eight pairs of $SiO_2$/$TiO_2$ layers (with thicknesses of 145 nm/100 nm, respectively), deposited by electron-beam deposition (eBeam) (Temescal Supersource) in vacuum, keeping the chamber at $10^{-5} \div 10^{-6}$ mbar throughout the process, at RT (deposition rates: 1 Å/s for $SiO_2$, 0.5 Å/s for $TiO_2$). The DBR is deposited on top of a 170 $\mu$m glass substrate and the resulting stopband is centered at 1476 meV (840 nm), as shown in Fig. 1A of the main text.

Single-layer $WS_2$ is mechanically exfoliated from bulk crystals with Nitto SPV 224 tape and transferred onto the surface of a Gel-Film (Gel-Pak WF x4 6.0 mil) substrate. Then single-layer flakes of $WS_2$ are identified on the PDMS stamp with an optical microscope operated in transmission mode and their thickness confirmed by differential reflectance. Single-layer WS$_2$ is transferred by all-dry deterministic transfer procedure on the DBR substrate by using a polydimethylsiloxane (PDMS) stamp. The sample is annealed in vacuum at 200 °C for 2h. Synthetic bulk WS$_2$ crystals are purchased from HQ Graphene.

After the evaporation of 20 nm-SiO$_2$ film via electron beam deposition on top of the WS$_2$ monolayer, gratings are spin-coated with PMMA 950k and a discharge layer; they are then positive-tone patterned by electron-beam-lithography at 30kV and finally developed.

**Optical Measurements** All measurements reported here are performed under ambient conditions at RT. The reflectivity measurements beyond the light cone are performed using an home-built microscope, equipped with a 1.42 numerical aperture (NA) oil immersion objective. Four lenses (Thorlabs, focal lengths of 20/30/50 cm) are used to project a magnified image of the back focal plane (BFP) onto the slit of an imaging spectrometer with a cooled charge-coupled device camera. For photoluminescence measurements, the WS$_2$ monolayer is off-resonant excited by using a continuous-wave 488 nm diode laser. The photoluminesce is recorded in reflection configuration, using a 40x objective with NA=0.95.

For nonlinear measurements, a tunable femtosecond laser (with pulse width $\sim$ 145 fs, repetition rate 10 kHz) is focused onto the BFP of the 0.95 NA objective. The energy of the laser resonantly pumping the interested polariton mode and the corresponding real space spot size dimensions are $\sim$ 1890 eV and $\sim$ 10 $\mu$m$^2$, respectively.

**Two-Coupled oscillator model** The experimental data reported in Fig. S7 and in Fig. 3 of the main text are fitted with a 2X2 coupled harmonic oscillator Hamiltonian:

$$H_{\mathbf{k}} = \begin{pmatrix} E_{ph}(k) & \Omega/2 \\ \Omega/2 & E_{exc} \end{pmatrix} \quad (1)$$

where $E_{exc}$ is the exciton energy, $\Omega$ is the Rabi splitting and $E_{ph}(k)$ is the energy dispersion of the BIC-like photonic branch, experimentally extrapolated from a PhC area without the TMD. We provide more details on the procedure in section E of the SI.

**Acknowledgements** The authors gratefully thank P. Cazzato for technical support and L. Dominici for fruitful discussion. The authors acknowledge the CINECA award under the ISCRA C initiative, for the availability of high performance computing resources and support.
This work was supported by the Italian Ministry of University (MIUR) for funding through the PRIN project "Interacting Photons in Polariton Circuits" — INPhoPOL (grant 2017P9FJBS), the project "Hardware implementation of a polariton neural network for neuromorphic computing"–Joint Bilateral Agreement CNR-RFBR (Russian Foundation for Basic Research) – Triennal Program 2021–2023, the MIUR project "ECOTEC - ECO-sustainable and intelligent fibers and fabrics for TEChnic clothing", PON « RI » 2014–2020, project N° ARS0100951, CUP B66C18000300005, the MAECI project "Novel photonic platform for neuromorphic computing", Joint Bilateral Project Italia-Polonia, 2022-2023 and the project "TECNOMED -Tecnopolo di Nanotecnologia e Fotonica per la Medicina di Precisione", (Ministry of University and Scientific Research (MIUR) Decreto Direttoriale n. 3449 del 4/12/2017, CUP B83B17000010001).

# Supplementary Information: Strongly enhanced light-matter coupling of a monolayer WS$_2$ from a bound state in the continuum

E. Maggiolini[1,2*], L. Polimeno[1*], F. Todisco [1], A. Di Renzo[1,3], M. De Giorgi[1], V. Ardizzone[1], R. Mastria[1], A. Cannavale[1,4], M. Pugliese[1], V. Maiorano[1], G. Gigli[1,3], D. Gerace[2], D. Sanvitto[1†], D. Ballarini[1†]

[1]*CNR NANOTEC, Institute of Nanotechnology, Via Monteroni, Lecce 73100, Italy*

[2]*Dipartimento di Fisica, Università di Pavia, Pavia, Italy*

[3]*Dipartimento di Matematica e Fisica E. De Giorgi, Università del Salento, Campus Ecotekne, Via Monteroni, Lecce 73100, Italy*

[4]*Department of Civil Engineering Sciences and Architecture, Polytechnic University of Bari, Bari, Italy*

† Corresponding author: dario.ballarini@nanotec.cnr.it, daniele.sanvitto@nanotec.cnr.it

## A    Optimal design concepts for BIC energy valleys embedded in quasi-single-mode synthetic microcavities

As already mentioned in the main text, we would like to meet a three-fold target, that is - in order of priority - maximizing the light-matter coupling, ensuring a BIC energy valley at a specific detuning from the exciton, and creating the conditions for a differential suppression of the lower-energy Bloch mode arising from the band-edge folding. First, we proceed to ensure the last two points by a wise choice of a general design concept, then meeting the upmost important target of the interaction enhancement by a metaheuristic parametric optimization over the structure of choice.

---

*These authors contributed equally: E. Maggiolini, L. Polimeno

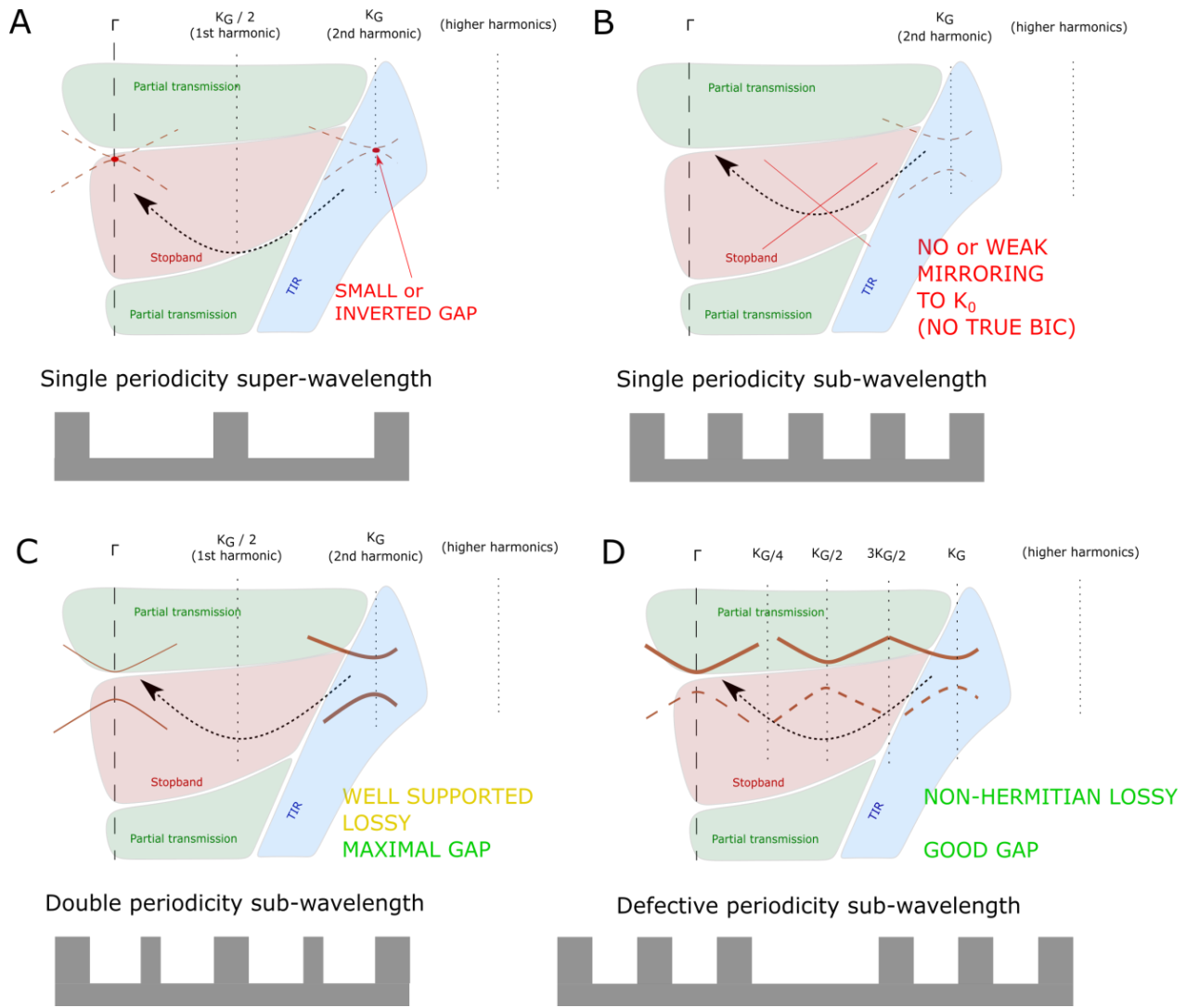

Figure S1: Comparison of the spectral content and consequent band-folding/mirroring phenomenology for different super and sub-wavelength $C_{2v}$-symmetry grating design solutions.

As a first point in the process, through qualitative considerations we can immediately see a number of clear advantages in a Bloch Surface Wave platform over a planar waveguide one: while a guided mode of a planar waveguide virtually possesses an infinite Q-factor due to the total internal reflection condition, this is mostly irrelevant in our case, since a Γ BIC solution is strictly unable to leak due to symmetry mismatch with respect to the continuum of radiation in free space. Of great relevance is instead the possibility to have a strong spatial enhancement of the in-plane electric field at a specific position across the vertical stack and the possibility to sustain modes within trenched and low-index parts of the grating structure; the first point is necessary for the interaction enhancement, but while in the BSW it can be attained with a simple and plain fabrication approach, in a waveguide it wouldn't be accessible; namely, it would require to include the TMD in the middle of patterned pillars for maximal field coupling, resulting in many disconnected and high edge-to-surface ratio pieces of active material; adding the possible additional damage introduced by etching, here a clear point in favour of BSWs can be found, irrespective of the grating design details. As we show in the leftmost column of Fig.S2a, a PhC slab on a simple superwavelength high index waveguide also fails to meet most of the other targets: starting from a small grating thickness of 90 nm *TiO₂* - in order to maximally enhance initial modal volume and mode overlap on the sides - the leftmost column in Fig.S2a clearly shows how this simple design leads to unsat-isfactory results: close to the 0.5 filling factor, which is the optimal target workpoint for the grating perturbativity, the x-averaged E-field maximum of the undesired Lossy branch optimally overlaps the TMD plane, while the E-field peak for the BIC branch remains always spatially shifted with respect to the active layer; at the same time, no effective filling factor can be found to obtain the BIC in an energy minimum. A solution to invert the BIC energy position - second column of Fig S2a - can be achieved by a partial etching, which leaves a limited high-index thickness everywhere; yet the energy gaps between the lossy and the BIC branch obtained by this solution are pretty small. Moreover they are obtained only in a small range of low filling factors, leaving small room for the tuning of the outcoupling leakage and visibility of the modes.

A remarkable improvement for both waveguides and BSW modes can be achieved going to subwavelength grating designs: it can be inferred from Fig.S1 that a high frequency component in the spatial Fourier spectrum of the grating is necessary to intersect the original confined mode beyond the light cone; the amplitude of the gaps for a subwavelength grating are irreproducible by a super-wavelength grating, for the simple reason in the latter the bandfolding is due to higher order harmonics of the Fourier spectrum, while in the subwavelength case the folding is due to the strong first harmonic.

The naive solution could be a simple subwavelngth grating, but as in Fig.S2 B, the problem here is the absence of any lower spectral component that is able to mirror the BSW folded beyond the light cone - around the X point in the reciprocal lattice - back to the Γ point at $k = 0$; even assuming to perform a small continuous breaking of the pillar symmetry (as in Fig.S1 c), the ef-fectiveness by which momentum is transferred between the X and Γ points is weak and depends on the amount of relative width difference between adiacent pillars; in other words, for this solution, the possibility to obtain a strongly accessible dispersion at $k = 0$ is thus intrinsically linked to the realization strong differences in the filling factors, requiring fabrication-inaccessible aspect ratios. A further disadvantage of design in Fig.S1 c would be the presence of a well supported Lossy mode, possibly competing with the BIC outside the protected Γ point in dispersion. Finally, the solution hereby adopted is shown in Fig S1d: we proof by numerical simulations that strongly perturbing the sub-wavelength grating structure by the subtraction of one pillar in each unit cell of the 1D lattice, allows to obtain a grating design with both a well defined and fabrication robust folding and mirroring components in the Fourier spectrum; as a consequence both the Lossy and BIC modes in principle increase their linewidth, but while in the BIC mode and closely around it in k-space this effect does not manifest, due to the topological symmetry-locked protection, in the lossy mode it does instead fully unfold, leading to a quasi-mode strongly interacting with the continuum; if interpreted in the framework of QNM (Quasi-Normal-Mode) theory[1], this introduces a greater degree of freedom in the numerical tuning of the dispersion characteristics: the strong non-hermiticity introduced for the lossy mode can be associated to a non-strictly-stationary nature of the modal field; in the language of QNM theory, quality factors and mode volumes will have to be generalized as complex quantities, losing their immediate physical meaning; nevertheless, a real-valued expression can still be derived for the Purcell enhancement of the emission rate:

$$\frac{\Gamma^{QNM}}{\Gamma_0} = \frac{\Gamma^{Hermitian}}{\Gamma_0} (1 + C_{N.H.}) \qquad (1)$$

where $\Gamma^{QNM}$ is the radiated power for a non-Hermitian system, described in terms of a Quasi-Normal-Modes basis, $\Omega$ is the energy of the mode and and $C_{N.H.}$ is a non-hermitian correction which can take both a positive or negative sign; if properly engineered though numerical optimization, a large and negative value for this correction can be leveraged to obtain a selective light-emitter decoupling for the undesired lossy mode. One could argue whether an optimal choice for introducing the defect in the subwavelenght lattice can be found, when considering our three-fold-targeted objective; the most sensitive objective here is the BIC position in the dispersion as an energy minimum or maximum: the answer is that choosing the remaining sub-wavelength units as an even or odd number, an unbalance towards a larger gap for a BIC-minimum or maximum can be found; both the BIC and lossy mode E-fields must be eigenstates of the $C_{2v}$ operator, due to the $C_{2v}$ point group symmetry of the potential (i.e. the dielectric profile in real space); in particular a $\Gamma - BIC$ spatial parity with respect to the symmetry axes is expected to be odd in order to be uncoupled from the continuum of radiation modes in free space, while the conjugate lossy one is expected to be even to be maximally coupled to it; a simple way to predict the relative energy position of the BIC and lossy modes is to look at their effective indices, mostly related to the weighted spatial distribution of the modes in the grating pillars (the higher the refractive index, the lower the mode energy) and in the grating trenches (the lower the index, the higher the mode energy); thus the even Lossy mode - in order not to be zero everywhere - is expected to have a symmetric and nonzero E-field at the symmetry axis of the unit cell, while the odd BIC mode is expected to have a node and a zero E-field at the same points; thus we can now clearly see how an odd number of pillars in each unit cell will result in a pillar on one of the two possible symmetry axes, where it will be differentially increasing the effective index of the Lossy mode, while it will remain mostly unweighted in the node of the BIC-like branch. As we show by numerical FDTD simulations in the rigthmost panels of FigS2 a and b, this solution allows to have a large energy gap opening between the lossy and the BIC branches, while having a robust possibility - e.g. in terms of filling factor tuning - of optimal BIC spatial coupling to the active material. We can see that that this solution applies here equally well to the waveguide and the BSW cases, but the BSW again is favoured due the more robust and TMD-safe fabrication workflow. This was thus our general design choice to perform fine automated parameter tuning on.

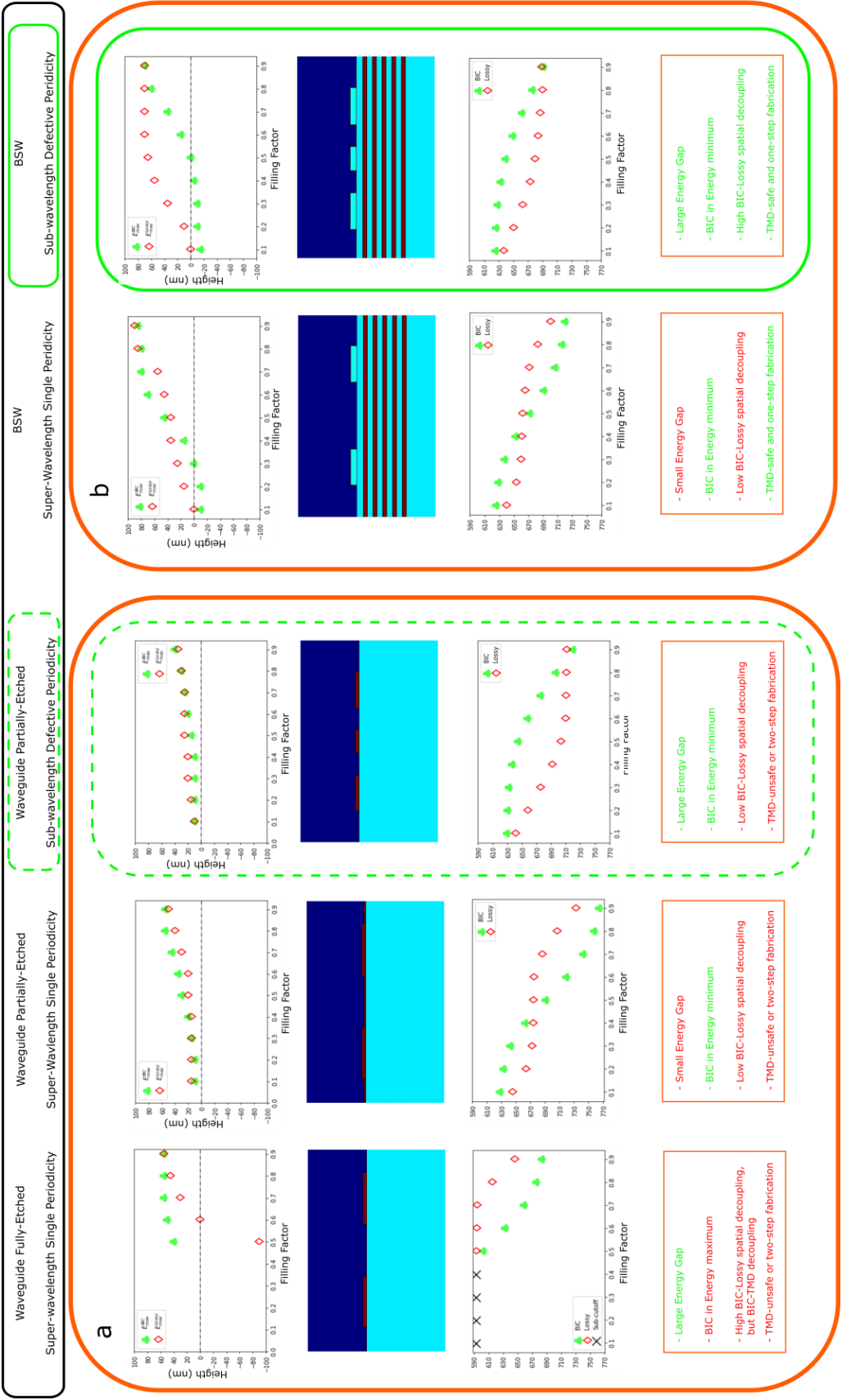

Figure S2: Comparison of different design solutions for the spatial and spectral separation of the Lossy mode from the BIC-bearing mode, with pros(green) and cons(red). In (a), moving from left to right, increasingly better solutions on a traditional high-index waveguide system, going from a simple fully-etched grating to a partially etched variant to a subwavelength defective grating. In (b) the same solutions for a BSW system, showing additional advantages for the subwavelength defect-bearing grating with respect to the same solution on a waveguide system.

## B  Metaheuristic optimization of the ligth-matter coupling in a BSW-TMD-PhC system

In order to efficiently guide an algorithmic search for the optimal light-matter coupling in this structure - while constraining to all our qualitative requirements on the dispersion characteristics - we adopt a metaheuristic optimization approach, embedding multiple physical considerations in an FDTD electromagnetic simulation setup, before the definition of a proper FOM (Figureof Merit) expression. The starting point is the parametrization of the structure to be optimized according to the qualitative considerations explained in Supplementary A; in particular, in this work we parametrize the defective 3-element grating unit with a periodicity and two different filling factors, one for the symmetry-axis-centered pillar and the other for two lateral pillars (see bottom-right grating concept in Fig. S1); in perspective, the possibilities offered by the algorithm can be increased by the addition of 2 further parameters for the glass spacer layer and the grating height, which we didn't introduce in this particular work, in order to use a previously pre-optimized fabrication process (i.e. we fixed the grating height to $\sim 170nm$) and a pre-fabricated DBR-WS2-SiO2 stack; as we will detail later, additional targets like the non-hermiticity-mediated suppression of the lossy branch, or additional $off - \Gamma$ accidental BICs could be enforced by the addition of these parameters to the optimization setting.

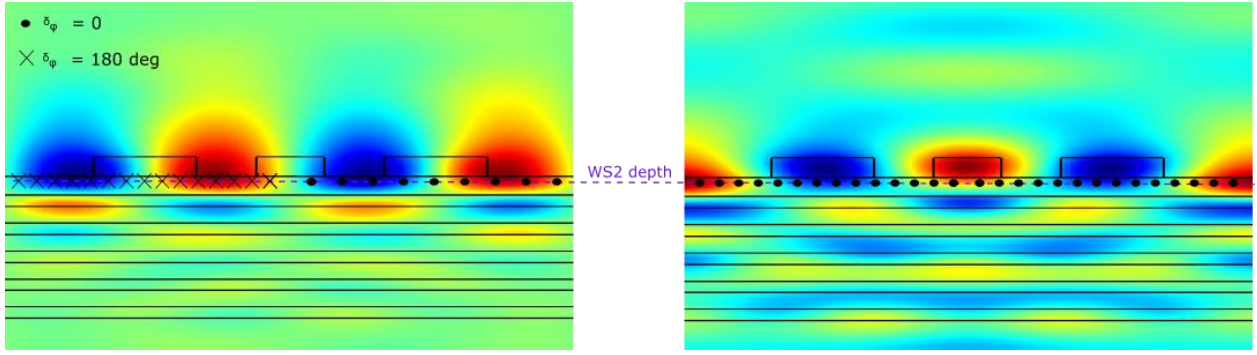

Figure S3: The different dipole generations across different iterations are symmetry-locked in phase and position with respect to the target dispersion branch to probe. Antysimmetric generations - as on the left - will selectively probe BIC-like dispersion branches, symmetric ones - on the right - will selectively probe the Lossy branch.

We need to stress that a global parametric optimization, even for a parametrization dimension chosen to be as small as 3 parameters, offers significant advantages when compared to brute force parameter sweeps, especially in the specific setting of a BSW system: the extreme sensitivity of a BSW mode to small changes in the dielectric environment at the air-DBR interface is a huge barrier to an effective optimization by parameter sweeps, since each change introduced by sweeping every single parameter (e.g. the grating period or filling factor), cannot be viewed in a perturbative picture on the basis of well defined unfolded modes; in general the virtual BSW mode folded by the grating periodicity will have phase and group effective indices that are largely perturbed by the averaged change introduced by the patterning; the linewidth will be even more heavily affected due to its energy dependence along the BSW mode dispersion; as a consequence, for instance, a specific BIC detuning can be obtained by virtually performing a demanding cross-sweep of the filling factor, the period, the spacer thickness and the grating height, but the resulting dispersion will not be equivalent, with different linewidths and group velocity dispersions for the folded modes; adding the Lossy mode non-hermitian behaviour to the picture clearly shows the complexity of this problem.

Due to the 1D nature of the grating and the large extension along the transverse direction to the periodicity, we can setup a FDTD electromagnetic simulation in a simple and computationally-economic 2D cell with Bloch boundary conditions and a dynamically-modulated and recentered spatial period; along the scattering direction, metallic boundary conditions are applied instead, cov- ered by Perfectly Matched Layers (PMLs). The illumination is provided by current-fixed broad- band electric dipole sources, uniformly placed in the plane of the TMD; the emission will thus depend on the radiation-impedance due to the confined-modes available to couple, thanks to the Purcell enhancement (i.e. Fermi Golden Rule in the limit of weak coupling); in each iteration, thedipoles are parametrized one by one with a continuously-variable phase, allowing to naturally em-bed them within our typical continuous variable search algorithm; thanks to the qualitative featuresenforced by the odd number of pillars in the symmetry-preserving defective grating unit, here weare able to introduce one important metaheuristic feature of our algorithm, that is we are able to enforce separate and selective probe simulations for the spatially-antisymmetric BIC branch and the spatially-symmetric lossy branch, respectively; in particular - as it is clear from Fig. S3 -, we force the target-mode symmetry in the dipole phases distribution around the symmetry axis of each test structure ; this means that we parametrize the dipole phase distribution with a num- ber of parameters corresponding to half the number of dipoles, by mirroring the pattern with a defined parity (+1 for the probing the Lossy branch, -1 for probing the BIC branch). The result of this approach is a symmetry-selective illumination of the system modes without the use of an un- physical and rigid selection of the symmetry within the boundary conditions; as a consequence, this approach - compared to common modal simulation approaches based for instance on Finite Differences in Frequency Domain - allows us to flexibly optimize on arbitrary points in the disper- sion, even beyond the $\Gamma$ point, where the symmetry is clear but weakly defined. For instance, this allows to co-optimize features such as additional accidental off-$\Gamma$ BICs, or make group velocity and linewidth engineering on a whole mode dispersion.

As to the FOM (Figure of Merit), here we use the radiated power in the top air far-field at the target BIC detuning from the exciton; this quantity is strictly related to the Purcell enhancement factor of the radiative emission from the dipoles, which - as we extensively discuss below - provides and indirect but reliable measure of the dipole-photon interaction strength both in weak coupling and within the strong-coupling regime as well; the choice to track the emission from the top side only (i.e. in the vacuum far-field), is the most reliable one, since the interpretation of the Poyinting vector projection in one direction, as related to the radiated power, can be robustly stated in the case of plane waves; the same can be not true in the near-field; for reasons of numerical convergence, we smooth the FOM manifold by integrating the emission within a finite frequency window rather than at a single energy, apodizing in any case the integration profile by a gaussian kernel; this other metaheuristic feature is of particular relevance in the search for a BIC, which due to the low-loss photonic environment allowed by our choice of materials ($SiO_2$, $TiO_2$, PMMA 950k), will result in a delta-like optimum, which would be impossible to map by any gradient-enhanced or gradient-free optimization algorithm.

For each iteration, test simulations are generated in batches corresponding to the number of parameters; each of these carries a different physical structure guess, a Bloch unit cell lenght and a symmetry axis dynamically repositioned accordingly, together with a different dipole-phase distribution guess; leveraging the continuous rather than discrete nature of the dipole phases, we are able to embed a sort of Montecarlo sweep of the illumination as part of a unique and smooth FOM manifold.

In the specific setting of this work, we only used the information from the BIC branch to define our FOM close to the Γ point; yet, more complex figures of merit can be build; In particular, by subtraction of a shadow FOM for the Lossy mode from the BIC FOM and leveraging on the strongly non-hermitian nature of the Lossy branch in our defective grating concept, a decoupling can in principle be attained for the Lossy branch with respect to the material dipoles; also, more general dispersion engineering can be done by our approach, e.g. inserting additional accidental BICs; this could be a precious tool in the quest for enhanced two-polariton parametric processes, obtained, e.g., by the engineering of high-DOS features in multiple points of the dispersion.

## C  Validation of a Figure of Merit from the weak to the strong coupling regime

We now discuss the validity of our classical simulation approach based on the chosen FOM, also to find the optimal conditions for the strong coupling regime: the Purcell enhancement factor of spontaneous emission is the notorious FOM to be optimized when analyzing light-matter coupling within the limit of the so called weak coupling regime. Such a quantity can be straigthforwardly defined as the ratio between the total radiative emission in a material or a resonant system and the radiative emission which the same emitter would produce in free space. Since our objective, as we mentioned before, is an accessible dispersion from the top air light-cone at a maximal light- matter interaction, we choose to optimize considering the far-field energy radiated to the air, ratherthan on the true Purcell factor - i.e. the total emission integrated around the dipoles. The two choices are not completely equivalent, first of all due to the radiation channel into the glass and the absorption from the materials; a topical point is that our chosen FOM can become a source ofconfusion, as the light-matter interaction gets strong enough to transition from a weak-coupling regime to a strong-coupling one: one can easily see that - due to disappearance of the original photonic modes and the emergence of new polaritonic eigenstates -, the mechanism of the Purcell enhancement could in principle easily break down, and it can become far from obvious that it produces a true enhancement of the emission; in weak coupling, the light-matter interaction can be seen as a perturbation on the separate excitonic and photonic states, and the exciton emission to thephotonic confined mode can be easily described by the Fermi golden rule; within such a picture, the emission enhancement with respect to free space is obvious, due to the increased density of statesfor the excitonic transition to occur, when energy/momentum conservation is satisfied. However, in strong coupling - as the perturbation gets comparable to the original eigenenergies - the picture breaks down and Fermi golden rule cannot be applied, with the eigenstates of the system branching away from the emitter in both momentum and energy.

In fact, as shown for example in ref. [2], as the light-matter coupling - quantified from the Rabi splitting - gets of the same order of magnitude (or beyond) the original eigenenergies, an actual suppression rather than enhancement of the radiative emission can be observed, due to quadratic terms in the vector potential, arising even within the minimal-coupling picture (i.e. in ∼ dipole interaction approximation). Hence - from a bootstrap extrapolation based on the experimental Rabi splittings and linewidths observed within our system - we would like to stress that a monotonous relation of the kind $\gamma \alpha \Omega^2$ well applies - with small corrections - in the transition from the weak to the strong coupling, up to light-matter interaction energies well beyond our workpoint.

Here we mostly follow the derivation in [2], which we refer for more details: we can approximately see our system as a dominating photonic harmonic oscillator (the BIC branch) coupled to an excitonic one; the model Hamiltonian can be written as:

$$H_{sys} = H_0 + H_{res} + H_{anti} + H_{el-bath} + H_{ph-bath} \quad (2)$$

In the limit of small excitation, both the exciton and photon oscillators can be assumed to obey bosonic commutation rules; in this context simple relations can be derived in input-output theory for the dependence of the radiative outcoupling from the interaction energy. The whole sys-tem Hamiltonian including the baths can then be modeled by boson-like creation and destruction operators, for which the single terms in eq. (2) can be represented in second quantization as C:

$$\begin{cases} H_0 = \sum_k \hbar\omega_{cav,k}\left(a_k^\dagger a_k + \frac{1}{2}\right) + \sum_k \hbar\omega_{12} b_k^\dagger b_k \\ H_{res} = \hbar \sum_k \left\{ i\Omega_{R,k}\left(a_k^\dagger b_k - a_k b_k^\dagger\right) + D_k\left(a_k^\dagger a_k + a_k a_k^\dagger\right) \right\} \\ H_{anti} = \hbar \sum_k \left\{ i\Omega_{R,k}\left(a_k b_{-k} - a_k^\dagger b_{-k}^\dagger\right) + D_k\left(a_k a_{-k} + a_k^\dagger a_{-k}^\dagger\right) \right\} \\ H_{el-bath} = \int dq \sum_k \hbar\omega_{q,k}^{el}\left(\beta_{q,k}^\dagger \beta_{q,k} + \frac{1}{2}\right) + i\hbar \int dq \sum_k \left(\kappa_{q,k}^{el} \beta_{q,k} b_k^\dagger - \kappa_{q,k}^{el*} \beta_{q,k}^\dagger b_k\right) \\ H_{ph-bath} = \int dq \sum_k \hbar\omega_{q,k}^{ph}\left(\alpha_{q,k}^\dagger \alpha_{q,k} + \frac{1}{2}\right) + i\hbar \int dq \sum_k \left(\kappa_{q,k}^{ph} \alpha_{q,k} a_k^\dagger - \kappa_{q,k}^{ph*} \alpha_{q,k}^\dagger a_k\right) \end{cases} \quad (3)$$

where $H_0$ is the term accounting for the uncoupled photonic and exciton harmonic oscillators (**a** and **b** operators respectively), $H_{res}$ and $H_{anti}$ are the resonant and antiresonant coupling terms between the two oscillators. The baths are linearly coupled to the system through coupling constants $\kappa$. One can then compute the dynamics of the various operators in the Heisenberg picture; without entering into the details of the calculation, the dynamics of the intracavity operators can be expressed as a sum of the evolution for the closed and unitary system, plus two contributions from the evolutions of the baths (last two terms in C):

$$\begin{cases} \frac{da_k}{dt} = -\frac{i}{\hbar}[a_k, H_{sys}] - \int_{t_0}^t dt' \Gamma_{ph,k}(t-t') a_k(t') + F_{ph,k}(t) \\ \frac{db_k}{dt} = -\frac{i}{\hbar}[b_k, H_{sys}] - \int_{t_0}^t dt' \Gamma_{exc,k}(t-t') b_k(t') + F_{el-exc,k}(t) \end{cases} \quad (4)$$

Extending the [$t_0$, $t$] interval to [$-\infty$, $\infty$] one can obtain Quantum Langevin equations for the system, when initial ("input") and final ("output") conditions for the bath creation operators can be pinned. Fourier transforming over the Heisenberg evolution equations, the whole dynamics of the system can then be cast into a matrix diagonalization problem, where the dissipative coupling to the bath - thanks to the convolution-like nature of the integrals above - becomes a simple diagonal entry in the $\overline{\mathcal{M}}_{k,\omega}$ matrix:

$$\overline{\mathcal{M}}_{k,\omega} \begin{pmatrix} \bar{a}_k(\omega) \\ \bar{b}_k(\omega) \\ \bar{a}^\dagger_{-k}(-\omega) \\ \bar{b}^\dagger_{-k}(-\omega) \end{pmatrix} + i \begin{pmatrix} \tilde{F}_{\text{ph},k}(\omega) \\ \tilde{F}_{el-exc,k}(\omega) \\ \tilde{F}^\dagger_{\text{ph},-k}(-\omega) \\ \tilde{F}^\dagger_{el-exc,-k}(-\omega) \end{pmatrix} = 0 \tag{5}$$

Where:

$$\overline{\mathcal{M}}_{k,\omega} = \begin{pmatrix} 2D - i\Gamma_{ph} - \omega + \omega_{ph} & i\Omega & 2D & -i\Omega \\ -i\Omega & -i\Gamma_{exc} - \omega + \omega_{exc} & -i\Omega & 0 \\ -2D & -i\Omega & -2D - \omega - \omega_{cav} - i\overline{\Gamma_{ph}} & i\Omega \\ -i\Omega & 0 & -i\Omega & -\omega - \omega_{exc} - i\overline{\Gamma_{exc}} \end{pmatrix} \tag{6}$$

Due to the assumed purely-dipole-like light-matter interaction in an exciton-polariton system, the D terms will take the form $D \sim \Omega^2/\omega_X$, leading to a typical Hopfield form for the inner-system Hamiltonian[3].

Now that the problem is well defined, we can perform a matrix inversion to write the inner-system creation operators as a function of the incoherent pumping from the bath, described by the Langevin forces $F_i$ (the dissipation to the bath is already embedded in the matrix $\overline{\mathcal{M}}_{k,\omega}$ by the $\Gamma$ terms). Pinning the initial and final conditions to the bath creation operators as $\alpha_{in},\alpha_{out},\beta_{in},\beta_{out}$ and solving for their evolution as well, simple algebraic relations linking the inner-system creationoperators and the bath initial and final conditions can be found; the solution for the inner-system creation operators can be eventually rewritten in terms of the input-output quantities; we can thenfinally yield relations linking the input and output for the whole open system:

$$\begin{pmatrix} \alpha^{out}_{q,\mathbf{k}} \\ \beta^{out}_{q',\mathbf{k}} \end{pmatrix} = \begin{pmatrix} \overline{\mathcal{U}}_{11}(\mathbf{k},\omega) & \overline{\mathcal{U}}_{12}(\mathbf{k},\omega) \\ \overline{\mathcal{U}}_{21}(\mathbf{k},\omega) & \overline{\mathcal{U}}_{22}(\mathbf{k},\omega) \end{pmatrix} \begin{pmatrix} \alpha^{in}_{q,\mathbf{k}} \\ \beta^{in}_{q',\mathbf{k}} \end{pmatrix} \quad (7)$$

We assume an initial nonresonant pumping, which can describe the non-resonant excitation in the experiment, as well as the wideband dipole-illumination that we employ in the FDTD simulations within the optimization process. This corresponds to pinning the expectation value for the photonic input bath number operator $<\alpha^{in\dagger}_{q,\mathbf{k}}|\alpha^{in}_{q,\mathbf{k}}>= 0$ and the material excitation bath $<\beta^{in\dagger}_{q,\mathbf{k}}|\beta^{in}_{q,\mathbf{k}}>\neq 0$

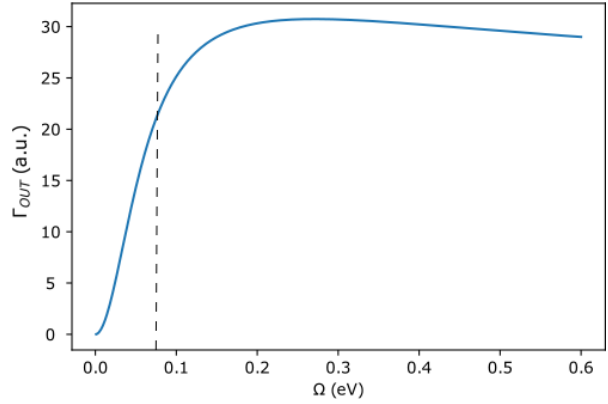

Figure S4: A) Radiation into free space as a function of Rabi coupling at the Rabi splitting point. The dashed line marks the Rabi splitting of our system, at around 70 meV.

We can now compute the output observable (photoluminescence), corresponding to the expectation value $<\alpha^{out}_{q,\mathbf{k}}|\alpha^{out}_{q,\mathbf{k}}>$:

$$<\alpha^{out}_{\mathbf{k}}|\alpha^{out}_{\mathbf{k}}> = \overline{\mathcal{U}}_{12}(\mathbf{k},\omega) = -2\,\mathrm{Re}\left\{\tilde{\Gamma}_{exc,\mathbf{k}}\right\} \frac{\kappa^{ph}_{\mathbf{k}}}{\kappa^{el-exc*}_{\mathbf{k}}}(\overline{\mathcal{M}}^{-1}_{\mathbf{k},\omega})_{1,2} \qquad (8)$$

The quantities $\kappa^{el-exc}_{\mathbf{k}}$ and $\tilde{\Gamma}_{exc,\mathbf{k}}$ are interrelated, with $\tilde{\Gamma}_{el-exc,\mathbf{k}} = \pi|\kappa^{el-exc}_{\mathbf{k}}|^2$, assuming a flat dispersion for the exciton. Here we give an order of magnitude for $\mathrm{Re}\{\tilde{\Gamma}_{exc,\mathbf{k}}\}$ coincident with the observed exciton PL linewidth; similarly, we take an estimation from the purely photonic dispersion of $\kappa^{ph}_{\mathbf{k}}$.

For each k, the true observed outcoupled radiation will be proportional to eq. 8 by the input free charge bath pumping and the density of states for the continuum, which we can assume to be fixed, having

$$\Gamma_{out} \sim \overline{\mathcal{U}}_{12}(\mathbf{k},\omega,\Omega) \qquad (9)$$

We then perform all the supermentioned derivations through symbolic computation tools[4], getting an analytical expression for Γ$_{out}$; in particular, we evaluate the expression at the anticrossing point for different values of the Rabi splitting, roughly fixing the linewidths by extrapolation from the experimentally observed PL:

$$\Gamma_{out}(\mathbf{k}(\omega = X - \Omega/2), \omega = X - \Omega/2, \Omega) \sim \overline{\mathcal{U}}_{12}(\mathbf{k}(\omega = X - \Omega/2), \omega = X - \Omega/2, \Omega) \quad (10)$$

The behavior we obtain is shown in Fig. S4, with the appearance of a non-monotonic deviation beyond our maximal interaction. As thoroughly described in ref. [5] using a planar cavity as a propotypical setting, this suppression of the radiative outcoupling is a fingerprint associated to an actual spatial decoupling of the light and matter components of the polariton in the near- field: the increase of the interaction energy Ω can have the paradoxical effect of reducing the true excitonic fraction beyond a certain point, so that despite the large interaction energy the observed polariton interactions can be eventually reduced. As we can see from Fig. S4, in our system we are close but still out of such a paradoxical regime of interaction - commonly defined Deep StrongCoupling; our choice to use a Purcell-like picture in the setting of our optimization algorithm is thus sound from this point of view. We would like to stress that the behaviour in Fig.S4 is derived at the Rabi Splitting point in our system; in any case we decide to optimize it at a point rather closer to the BIC than to the anticrossing point; as a consequence and for the sake of clarity, the behaviour in Fig. S4 should also be considered a worst case scenario, with respect to the true interaction involved in the optimization process of the BIC at some larger detuning than the Rabi splitting.

## D    BSW-WS2 oil-immersion characterization

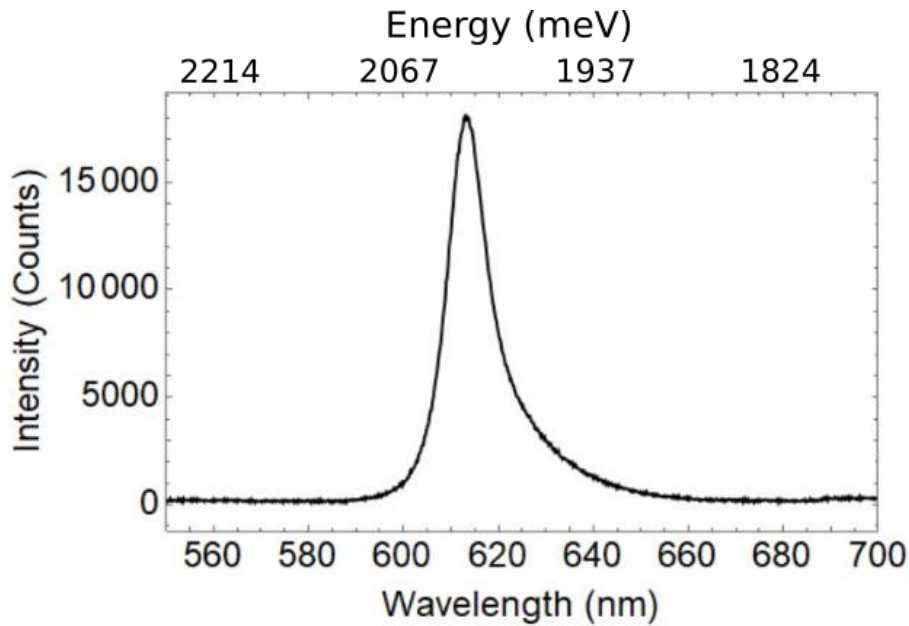

Figure S5: Monolayer photoluminescence spectrum under 2541 meV/488 nm CW-excitation. The single peak is centred at 2016 meV with a FWHM of ∼ 30 meV.

A $WS_2$ monolayer is transferred on top of a DBR with a stopband centered at 1476 meV. Figure S5 shows the photoluminescence spectrum obtained by pumping the material off resonance, with 2541 meV/488 nm-CW laser.

Figure S6A shows the experimental dispersion of the bare Bloch surface mode supported by the dielectric mirror. The experimental data are collected with an oil immersion objective in order to detect the signal from the total internal reflection region. Figure S6B shows the experimental dispersion of the upper and lower polariton modes measured from reflectivity with white light in the centre of the monolayer. Their anti-crossing around the exciton energy at 2016 meV is evident

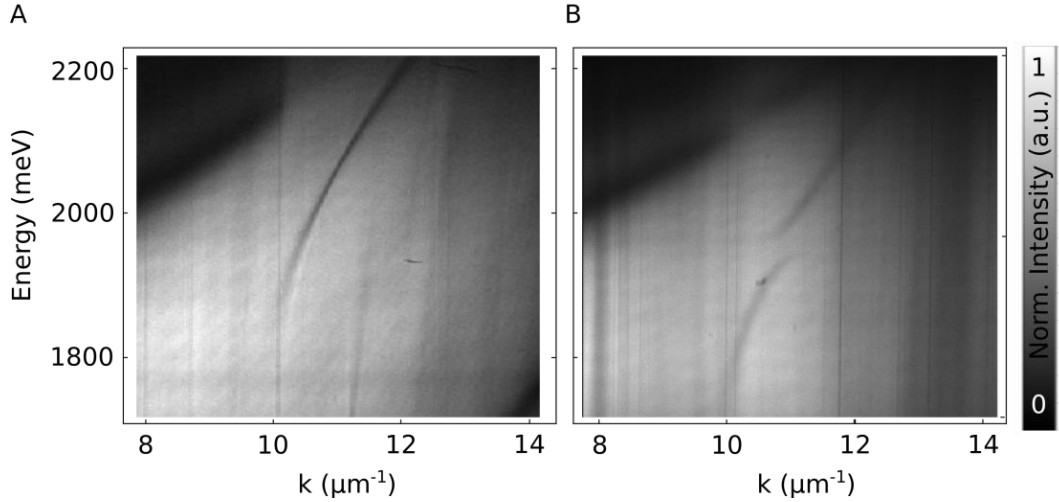

Figure S6: A) Experimental dispersion of the bare Bloch surface mode. B) Experimental dispersion of BSW polaritons in monolayer $WS_2$ in reflectance.

in the reflectance.

## E  BIC-polariton Rabi-splitting characterization method

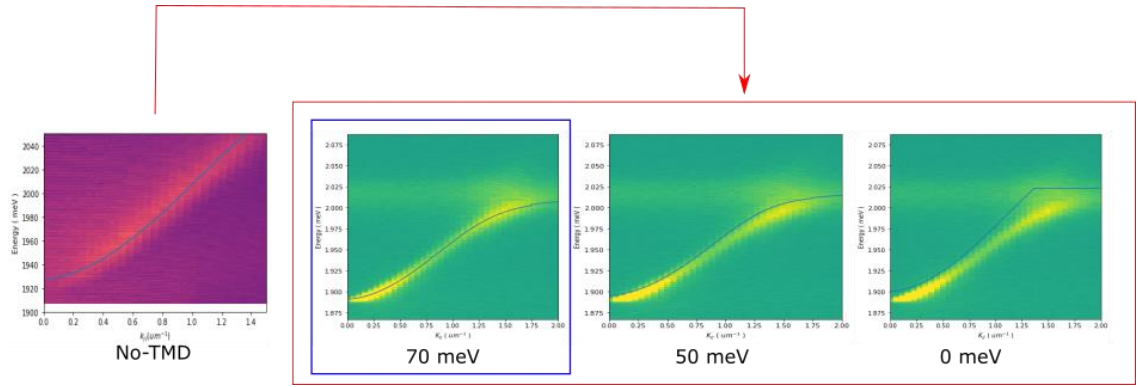

Figure S7: A) Fitting for the Rabi splitting in our BIC-BSW-TMD system, assuming a constant energy shift of 27 meV retrieved from an FDTD simulation and accounting for the TMD average dielectric response (no resonance).

From a general point of view, we would like to derive the Rabi splitting by fitting the disper-sion with the eigenvalues of the matrix:

$$H_{\mathbf{k}} = \begin{pmatrix} E_{ph}(k) & \Omega/2 \\ \Omega/2 & E_{exc} \end{pmatrix} \quad (11)$$

Nevertheless, we have a subtle issue arising in our system: as one can easily see in Fig.S6, the curvature of the BSW mode is clearly non-linear; the same is expected to happen on the folded dispersion emerging after the insertion of the PhC structure and spacer. Correctly fitting the $H_k$ matrix at the true crossing point between the exciton resonance and the photonic BIC-like mode is a difficult task to perform on the basis of the strongly-coupled dispersion only, since the true photonic dispersion $E_{BIC\text{-}branch}$ will be an unknown, as well as the Rabi Splitting $\Omega$. In order to improve the accuracy of our evaluation, we proceed to extract the true photonic dispersion from a grating area without the WS2 ML embedded: as a first point, we finely recalibrate the dispersion of the PL data to fit to the reference photonic dispersion Lossy branch, which we do not expect to change due to the lower coupling and large detuning to the exciton. In fact, we do not even observe any shift in the peak of the mode at the Γ point. We then proceed to fit the photonic BIC-like dispersion with a catenary function of a 7-order polynomial in k (Fig.S7 leftmost). We then superimpose the analitically defined function to the PL dispersion; a subtle point to consider here is the average dielectric response of the TMD, which is separate from the neutral exciton resonance; we make an estimation of this contribution to the BIC-branch in the order of $\sim 25meV$, adding it as a shift to the original photonic dispersion function.

As we show in Fig.S7, the Rabi splitting can be estimated in the order of $\sim 70meV$, assuming the correct neutral exciton position as the one cleanly extracted on the bare BSW stack.